\documentclass[12pt,english]{paper}
\usepackage{amssymb}
\usepackage[T1]{fontenc}
\usepackage[latin1]{inputenc}
\usepackage{graphicx}
\usepackage{cite}
\makeatletter

\newcommand{\bibstyle@aas}{\bibpunct{(}{)}{;}{a}{}{,}}
\makeatletter

\makeatletter

\usepackage{geometry}

\makeatletter

\usepackage{epsfig}

\makeatother

\makeatother

\makeatother

\usepackage{babel}
\makeatother
\begin{document}

\title{Measurement of TeV dark particles due to decay of heavy dark matter in the earth core at IceCube}

\author{Ye Xu$^{1,2}$}

\maketitle

\begin{flushleft}
$^1$School of Information Science and Engineering,  Fujian University of Technology, Fuzhou 350118, China
\par
$^2$Research center for Microelectronics Technology, Fujian University of Technology, Fuzhou 350118, China
\par
e-mail address: xuy@fjut.edu.cn
\end{flushleft}

\begin{abstract}
In the present paper, it is assumed that there exist two species of dark matter: a heavy dark matter particle (HDM) with the mass of O(TeV) which is generated in early universe and a lighter dark matter particle (LDM) which is a relativistic product due to the decay of HDM. HDMs, captured by the earth, decay to high energy LDMs, and these particles can be measured by km$^3$ neutrino telescopes, like the IceCube detector. A $Z^{\prime}$ portal dark matter model is taken for LDMs to interact with nuclei via a neutral current interaction mediated by a heavy gauge boson $Z^{\prime}$. With the different lifetimes of decay of HDMs and Z$^{\prime}$ masses, the event rates of LDMs, measured by IceCube, are evaluated in the energy range between 1 TeV and 100 TeV. According to the IceCube data, the upper limit for LDM fluxes is estimated at 90\% C.L. at IceCube. Finally, it is proved that LDMs could be directly detected in the energy range betwen O(1TeV) and O(10TeV) at IceCube with $m_{Z^{\prime}} \lesssim 500 GeV$ and $\tau_{\phi} \lesssim 10^{21}$ s.
\end{abstract}


\section{Introduction}
The nature and origin of dark matter (DM) remains one of the unanswered puzzles in particle physics, cosmology and astrophysics, but sufficient evidences for the existence of DM and its dominance in matter in our universe are provided by cosmological and astrophysical observations\cite{bergstrom,BHS}. $26.6\%$ of the overall energy density of the universe is nonbaryonic DM\cite{Planck2015}. Weakly Interacting Massive Particles (WIMPs), predicted by extensions of the Standard Model (SM) of particle physics, are a class of candidates for dark matter\cite{BHS} and distributed in a halo surrounding a galaxy. This WIMP halo with a local density of 0.3 GeV/cm$^3$ is assumed and its relative speed to the Sun is 230 km/s\cite{JP}. At present, one mainly searches for WIMPs via direct and indirect detections\cite{CDMSII,CDEX,XENON1T,LUX,PANDAX,AMS-02,DAMPE,fermi}. Because of the very small cross sections of the interactions between these WIMPs and nuclei\cite{XENON1T,PANDAX}, so far no one has found this thermal DM yet.
\par
A heavy dark matter particle (HDM) $\phi$ is an alternative DM scenario\cite{KC87,CKR98,CKR99,KT,KCR,CKRT,CCKR,KST,CGIT,FKMY,FKM}. Then it is a reasonable assumption that there exist two DM species in the Universe. One is a non-relativistic dark sector generated by the early universe with its bulk comprised of a massive relic ($m_{\phi} \sim O(TeV)$) in the Universe. The other is a stable lighter dark matter particle (LDM) $\chi$ which is products of the decay of HDMs ($\phi\to\chi\bar{\chi}$). It is assumed that HDMs comprise of the bulk of present-day DM. Since the decay of long-living HDMs ($\tau_{\phi}$ > $t_0$\cite{AMO,EIP}, $t_0$ is the age of the universe), meanwhile, the present-day DM may also contain a small component which is high energy LDMs. Besides direct measurements of HDMs, one can detect products of the decay of HDMs. That is a flux of protons, gamma rays and neutrinos is produced by the decay of HDMs into the partons. Conversely, it is different from this decay mode referred to above that the products of the decay of HDMs are a class of lighter fermion dark matter\cite{BGG,BGGM,xu1,xu2,xu3}, not SM particles, in the present work. Although the fraction of these relativistic LDM particles is small in the Universe, their large interaction cross sections (including between themselves and between them and SM particles) make it possible to measure them. In the present paper, a Z$^{\prime}$ portal dark matter model\cite{APQ,Hooper} is taken for LDMs $\chi$ to interact with nuclei. And, for the $\chi\chi$Z$^{\prime}$ and qqZ$^{\prime}$ interactions, their vertexes are both assumed to be vector-like. These HE LDM particles may be directly measured by their interaction with nuclei. Thus it is indicated that there exist HDMs in the Universe.
\par
HDMs, captured by the earth, decay to high energy LDMs ($\phi\to\chi\bar{\chi}$), and these particles ($\chi$), which pass through the Earth and ice and interact with nuclei, can be measured by km$^3$ neutrino telescopes, such IceCube\cite{icecube2017}, ANTARES\cite{antares}, KM3NeT\cite{km3net}. In this measurement, the main contamination is from astrophysical and atmospheric neutrinos. In what follows, the event rates of LDMs, measured by IceCube, from the earth core will be evaluated in the energy range between 1 TeV and 100 TeV with the different lifetimes of decay of HDMs and Z$^{\prime}$ masses. Thus it will be proved that the possibility of measurement of these TeV LDM at IceCube.
\par
\section{HDM accumulation in the Earth}
It is considered a scenario where the dark matter sector is composed of two particle species in the Universe. One is a non-relativistic particle species $\phi$, with mass of O(TeV), that is between 1 TeV and 100 TeV in the present paper, the other is much lighter particle species $\chi$ ($m_{\chi} \ll m_{\phi}$), due to the decay of $\phi$, with a very large lifetime. And $\phi$ comprises the bulk of present-day DM. The lifetime for the decay of HDMs to SM particles is strongly constrained ($\tau \geq$ O($10^{26}-10^{29}$)s) by diffuse gamma and neutrino observations\cite{EIP,MB,RKP,KKK}. However, since in this scenario $\phi$ does not decay to SM particles, constraints relevant here are only those based on cosmology. Then, in the present work, it is considered an assumption that HDMs only decay to LDMs. So $\tau_{\phi}$ is taken to be between $10^{17}$ s (the age of the Universe) and $10^{26}$ s.
\par
When the HDM wind sweeps through the Earth, those non-relativistic particles could collide with the matter in the Earth and lose their kinetic energy. Then they can be captured by the Earth's gravity and enter the earth core. After a long period of accumulation, the HDMs inside the Earth can begin to decay into LDMs at an appreciable rate. The number N of HDMs, captured by the Earth, obeys the equation\cite{BCH}
\par
\begin{center}
\begin{equation}
\frac{dN}{dt}=C_{cap}-2\Gamma_{ann}-C_{evp}N
\end{equation}
\end{center}
\par
where $C_{cap}$ is the capture rate, $\Gamma_{ann}$ is the annihilation rate and $C_{evp}$ is the evaporation rate. The annihilation of HDMs can be ignored since their cross section is bounded by unitarity ($\sigma_{ann} \propto \displaystyle\frac{1}{m_{\phi}^2}$). The evaporation rate is only relevant when the DM mass < 5 GeV\cite{gould,KSW,nauenberg,GS}, which are much lower than our interested mass scale (m$_{\phi}$ $\geq$ 1 TeV). Thus the HDM evaporation is ignored in my work.
\par
$C_{cap}$ is proportional to $\displaystyle\frac{\sigma_{\phi N}}{m_{\phi}}$\cite{JKK} and  $\sigma_{\phi N}$ is the scattering cross section between the nucleons and HDMs. The $C_{cap}$ calculation, with the Gaussian-no solar depletion model in Ref.\cite{LE}, is adopted but $\sigma_{\phi N}$ is taken to be 10$^{-44}$ cm$^2$ for $m_{\phi} \sim$ O(TeV) \cite{XENON1T,PANDAX}.
\section{LDM and neutrino interactions with nuclei}
In the present paper, a $Z^{\prime}$ portal dark matter model\cite{APQ,Hooper} is taken for LDMs to interact with nuclei via a neutral current interaction mediated by a heavy gauge boson $Z^{\prime}$. This new boson is considered as a simple and well-motivated extension of SM (see Fig.1(a) in Ref.\cite{BGG}). Since the interaction vertexes ($\chi\chi Z^{\prime}$ and $qqZ^{\prime}$) are assumed to be vector-like in the present work, the effective interaction Lagrangian can be written as follows:
\begin{center}
\begin{equation}
\mathcal{L} = \bar{\chi}g_{\chi\chi Z^{\prime}}\gamma^{\mu}\chi Z^{\prime}_{\mu} + \sum_{q_i} \bar{q_i}g_{qqZ^{\prime}}\gamma^{\mu}q_iZ^{\prime}_{\mu}
\end{equation}
\end{center}
where $q_i$'s are denoting the SM quarks, and $g_{\chi\chi Z^{\prime}}$ and $g_{qqZ^{\prime}}$ are denoting the $Z^{\prime}$-$\chi$ and $Z^{\prime}$-$q_i$ couplings, respectively.
This deep inelastic scattering (DIS) cross-section is computed in the lab-frame. The coupling constant G ($G=g_{\chi\chi Z^{\prime}}g_{qqZ^{\prime}}$) is chosen to be 0.05 and the $\chi$ masse is taken to be 10 GeV, and the $Z^{\prime}$ mass is taken to be 100 GeV, 250 GeV and 500 GeV, respectively. Theoretical models that encompass the LDM spectrum have been discussed in the literature in terms of Z or $Z^{\prime}$ portal sectors with $Z^{\prime}$ vector boson typically acquiring mass through the breaking of an additional U(1) gauge group at the high energies (see Ref.\cite{APQ,Hooper}). The DIS cross section for LDM interaction is given by a simple power-lay form\cite{BGG}. With $m_{Z^{\prime}} = 100 GeV$, its cross section is obtained by the following function:
\begin{center}
\begin{equation}
\sigma_{\chi N}=3.83\times10^{-36} cm^2 \left(\frac{E_{\chi}}{1GeV}\right)^{0.518}
\end{equation}
\end{center}
where E$_{\chi}$ is the LDM energy.
\par
The DIS cross-sction for neutrino interaction with nuclei is computed in the lab-frame and given by simple power-law forms\cite{BHM} for neutrino energies above 1 TeV:
\begin{center}
\begin{equation}
\sigma_{\nu N}(CC)=4.74\times10^{-35} cm^2 \left(\frac{E_{\nu}}{1 GeV}\right)^{0.251}
\end{equation}
\end{center}
\par
\begin{center}
\begin{equation}
\sigma_{\nu N}(NC)=1.80\times10^{-35} cm^2 \left(\frac{E_{\nu}}{1 GeV}\right)^{0.256}
\end{equation}
\end{center}
where $\sigma_{\nu N}(CC)$ and $\sigma_{\nu N}(NC)$ are the DIS cross-sections for neutrino interaction with nuclei via a charge current (CC) and neutral current (NC), respectively. $E_{\nu}$ is the neutrino energy.
\par
The LDM and neutrino interaction lengths can be obtained by
\begin{center}
\begin{equation}
L_{\nu,\chi}=\frac{1}{N_A\rho\sigma_{\nu,\chi N}}
\end{equation}
\end{center}
\par
where $N_A$ is the Avogadro constant, and $\rho$ is the density of matter, which LDM and neutrinos interact with.
\section{Evaluation of the numbers of LDM and neutrinos measured by IceCube}
IceCube is a km$^3$ neutrino telescope and can detect three flavour neutrinos via detecting the secondary particles, that in turn emit Cherenkov photons, produced by the interaction between neutrinos and the Antarctic ice\cite{icecube2018}. HDMs, captured by the earth, decays to high energy LDMs, and these particles from the earth core pass through the Earth and ice. Meanwhile they interact with matter of the Earth and ice. Cherenkov Photons are produced by cascades due to LDM interaction with nuclei within the IceCube (see Fig. 1). A small part of these photons will be detected by the IceCube detector. Since these LDMs interact with the nuclei in the ice and this is very similar to DIS of neutrino interaction with nuclei via a neutral current, its secondary particles develop into a cascade at IceCube. In the present paper, it is assumed that all LDMs and neutrinos, detected by IceCube, interact with the ice within its volume.
\par
The number of LDMs, due to the decay of HDMs in the earth core in ten years, N$_{decay}$ can be obtained by the following function:
\begin{center}
\begin{equation}
N_{decay} = 2N_0\left(exp(-\frac{t_0}{\tau_{\phi}})-exp(-\frac{t_0+T}{\tau_{\phi}})\right)
\end{equation}
\end{center}
where $N_0=\int^{t_e}_0 C_{cap} dt$ is the number of HDMs captured by earth, t$_0$ is the age of the universe and T is the lifetime of taking data for IceCube and taken to be 10 years in the present work. t$_e$ is the age of the earth.
\par
The number of LDMs, detected by IceCube, N$_{det}$ can be obtained by the following function:
\begin{center}
\begin{equation}
N_{det} = R\times T\times \int^{E_{max}}_{E_{min}} \int_{S_{eff}} \eta \frac{N_{decay}}{4\pi R_e^2} P(E) dS dE
\end{equation}
\end{center}
where R is the duty cycle for IceCube and taken to be 100\%. dS is the surface element. $S_{eff}$ is the effective observational area for IceCube and about 1 km$^2$. E is the energy of an incoming particle and varies from $E_{min}$ to $E_{max}$. R$_e$ is the radius of the Earth. $P(E)=exp(-\displaystyle\frac{D_{earth}}{L_{earth}})\left(1-exp(-\displaystyle\frac{D}{L_{ice}})\right)$. $L_{earth,ice}$ is the LDM interaction lengths with the earth and ice, respectively. D is the effective length in the IceCube detector. Since $N_{det}$ is just Roughly estimated in the present paper, the efficiency for measuring showers produced by LDMs or neutrinos, $\eta$, is assumed to be 100\%.
\par
The background is mainly two sources: astrophysical and atmospheric neutrinos which pass through the earth core and reach IceCube. The astrophysical neutrinos are estimated with a diffuse neutrino flux of $\Phi_{\nu}^{astro}=0.9^{+0.30}_{-0.27}\times\left(\displaystyle\frac{E_{\nu}}{100TeV}\right)^{-2.13\pm0.13}\times10^{-18}GeV^{-1} cm^{-2}s^{-1}sr^{-1}$\cite{icecube2016}, where $\Phi_{\nu}^{astro}$ represents the per-flavor flux, by the above method. And the atmospheric neutrinos is estimated with a flux of $\Phi_{\nu}^{atm} = C_{\nu}\left(\displaystyle\frac{E_{\nu}}{1GeV}\right)^{-(\gamma_0+\gamma_1y+\gamma_2y^2)}$ by the same method\cite{SMS}, where $\Phi_{\nu}^{atm}$ represents the atmospheric neutrino flux, $y=log_{10}(E_{\nu}/1GeV)$, and the coefficients, $C_{\nu}$ ($\gamma_0$, $\gamma_1$ and $\gamma_2$) are given in Table III in Ref.\cite{SMS}.
\section{Results}
The event rates of LDMs and neutrinos, detected by IceCube, are evaluated in the energy between 1 TeV and 100 TeV, respectively.  The numbers of the detected LDMs can reach about 1974 and 1 at 1 TeV and 47 TeV with $m_{Z^{\prime}}=100 GeV$ and $\tau_{\phi} = 10^{18}$ s in ten years, respectively, as shown in Fig. 2 (see the red solid line). The numbers of the detected LDMs can reach about 67 and 1 at 1 TeV and 16 TeV with $m_{Z^{\prime}}=250 GeV$ and $\tau_{\phi} = 10^{18}$ s in ten years, respectively, as shown in Fig. 2 (see the blue solid line). The numbers of the detected LDMs can reach about 4 and 1 at 1 TeV and 2.4 TeV with $m_{Z^{\prime}}=500 GeV$ and $\tau_{\phi} = 10^{18}$ s in ten years, respectively, as shown in Fig. 2 (see the purple solid line). The numbers of the detected LDMs can reach about 1974 and 1 at 1 TeV and 47 TeV with $m_{Z^{\prime}}=100 GeV$ and $\tau_{\phi} = 10^{18}$ s in ten years, respectively, as shown in Fig. 3 (see the red solid line). The numbers of the detected LDMs can reach about 291 and 1 at 1 TeV and 21 TeV with $m_{Z^{\prime}}=100 GeV$ and $\tau_{\phi} = 10^{19}$ s in ten years, respectively, as shown in Fig. 3 (see the green solid line). The numbers of the detected LDMs can reach about 30 and 1 at 1 TeV and 7 TeV with $m_{Z^{\prime}}=100 GeV$ and $\tau_{\phi} = 10^{20}$ s in ten years, respectively, as shown in Fig. 3 (see the blue solid line). The numbers of the detected LDMs can reach about 3 and 1 at 1 TeV and 2 TeV with $m_{Z^{\prime}}=100 GeV$ and $\tau_{\phi} = 10^{21}$ s in ten years, respectively, as shown in Fig. 3 (see the purple solid line).
\par
The event rate of neutrinos is at least smaller by 5 orders of magnitude within the energy range where LDMs can be measured at IceCube, compared to that of LDMs, as shown in Fig. 2 and 3. So the neutrino contamination is neglected at all.
\section{Conclusion}
According to the IceCube data from 2008 to 2015\cite{icecubedata2017}, no events from the earth core are measured at IceCube in this period of time. So the upper limit for the number of LDMs $N_{up}$ is equal to 2.44 at 90\% C.L. with the Feldman-Cousins approach\cite{FC}. Fig. 4 shows LDM fluxes estimated with $m_{Z^{\prime}}=100 GeV$ s (red solid line), $250 GeV$ (blue solid line) and $500 GeV$ (purple solid line) and the upper limit for LDM flux at 90\% C.L. (black solid line) at IceCube. This limit excludes LDM flux below 13 TeV (with $m_{Z^{\prime}} = 100 GeV$), 5 TeV (with $m_{Z^{\prime}} = 250 GeV$) and 1.8 TeV (with $m_{Z^{\prime}} = 500 GeV$), respectively. So LDMs, due to the decay of HDMs in the earth core, could be probed in the energy range between 13 TeV and 47 TeV (with $m_{Z^{\prime}}=100 GeV$ and $\tau_{\phi} = 10^{18}$ s), 4.3 TeV and 16 TeV (with $m_{Z^{\prime}}=250 GeV$ and $\tau_{\phi} = 10^{18}$ s) and 1.5 TeV and 2.4 TeV (with $m_{Z^{\prime}}=500 GeV$ and $\tau_{\phi} = 10^{18}$ s) at IceCube in ten years, respectively.
\par
According to the results described above, it is possible that LDMs are directly detected in the energy range between O(1TeV) and O(10TeV) at IceCube with $m_{Z^{\prime}} \lesssim 500 GeV$ and $\tau_{\phi} \lesssim 10^{21}$ s. Thus This might prove whether there exist HDMs in the Universe. Certainly, the above results are based on the assumption that the efficiencies for measuring LDMs and neutrinos are set to be 100\%. Besides, sufficient exposure will be used to determine whether it is possible that LDMs are detected at a km$^3$ neutrino telescope.
\section{Acknowledgements}
This work was supported by the National Natural Science Foundation
of China (NSFC) under the contract No. 11235006, the Science Fund of
Fujian University of Technology under the contracts No. GY-Z14061 and GY-Z13114 and the Natural Science Foundation of
Fujian Province in China under the contract No. 2015J01577.
\par

\newpage

\begin{figure}
 \centering
 \includegraphics[width=0.9\textwidth]{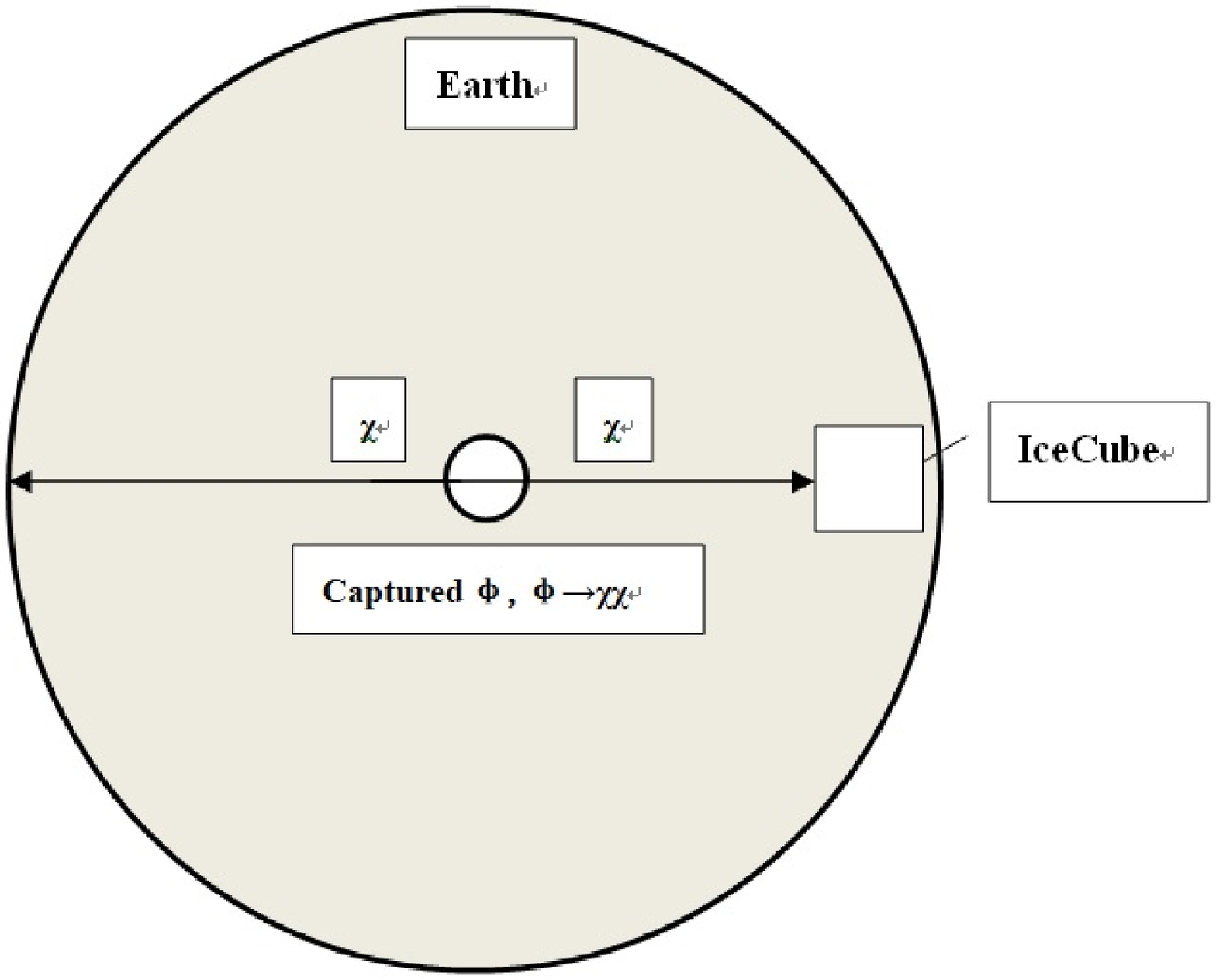}
%%% bb = left_bottom_X, left_bottom_Y, right_top_X, right_top_Y
%%% scale through "set width"
 \caption{LDMs, due to the decay of HDMs captured in the earth core, pass through the Earth and ice and can be measured by IceCube}
 \label{fig:fig}
\end{figure}

\begin{figure}
 \centering
 \includegraphics[width=0.9\textwidth]{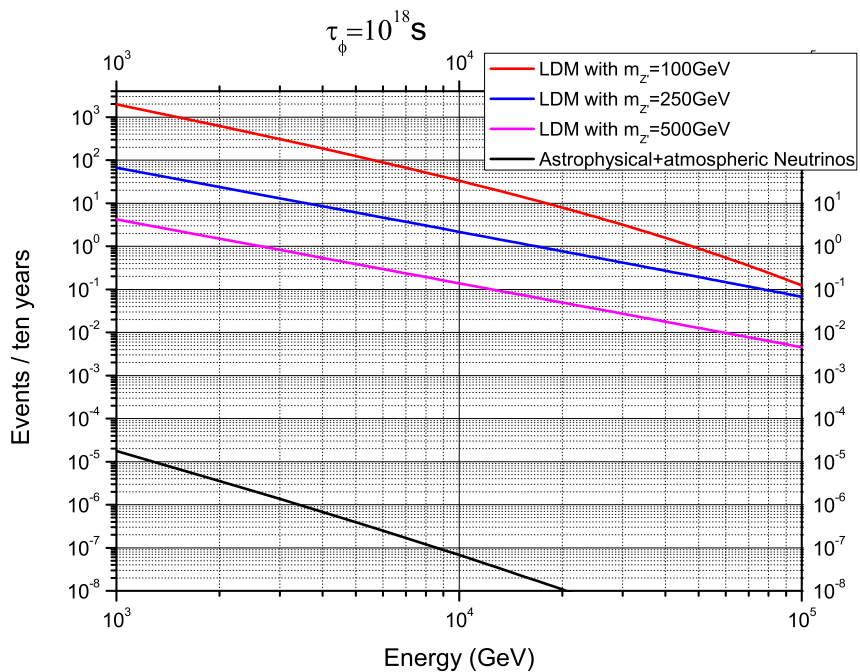}
%%% bb = left_bottom_X, left_bottom_Y, right_top_X, right_top_Y
%%% scale through "set width"
 \caption{ The numbers of the detected LDMs and neutrinos are evaluated at IceCube with the different $Z^{\prime}$ masses (100 GeV, 250 GeV and 500 GeV) and $\tau_{\phi} = 10^{18}$ s in ten years, respectively}
 \label{fig:results1}
\end{figure}

\begin{figure}
 \centering
 %\includegraphics[bb=0 0 200 300, width=3.8cm]{energy}
 %\hspace{0.7\textwidth}
 \includegraphics[width=0.9\textwidth]{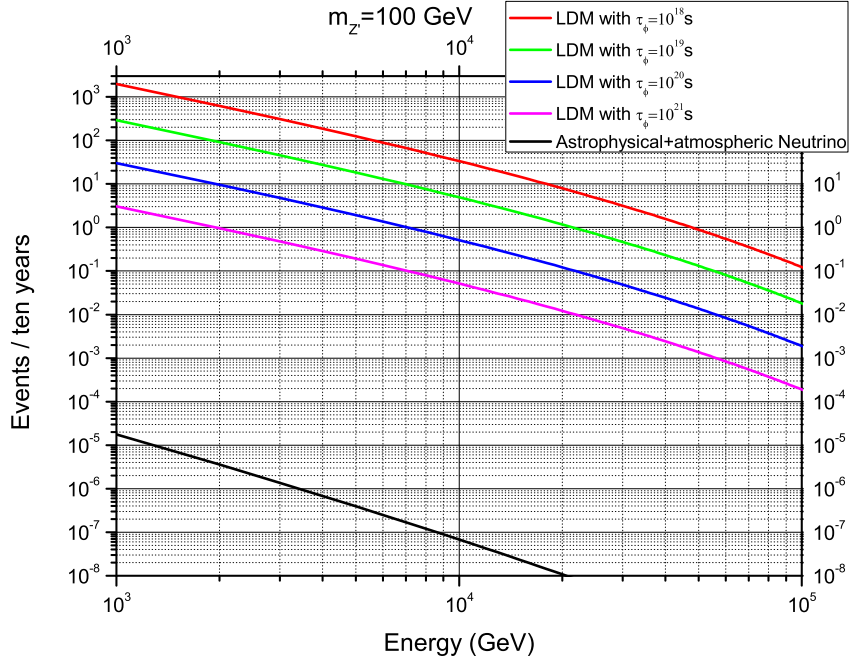}
 \caption{The numbers of the detected LDMs and neutrinos are evaluated at IceCube with the different $\tau_{\phi}$ ($10^{18}$s, $10^{19}$s, $10^{20}$s and $10^{21}$s) and $m_{Z^{\prime}}=100 GeV$ in ten years, respectively}
 \label{fig:results2}
\end{figure}

\begin{figure}
 \centering
 %\includegraphics[bb=0 0 200 300, width=3.8cm]{energy}
 %\hspace{0.7\textwidth}
 \includegraphics[width=0.9\textwidth]{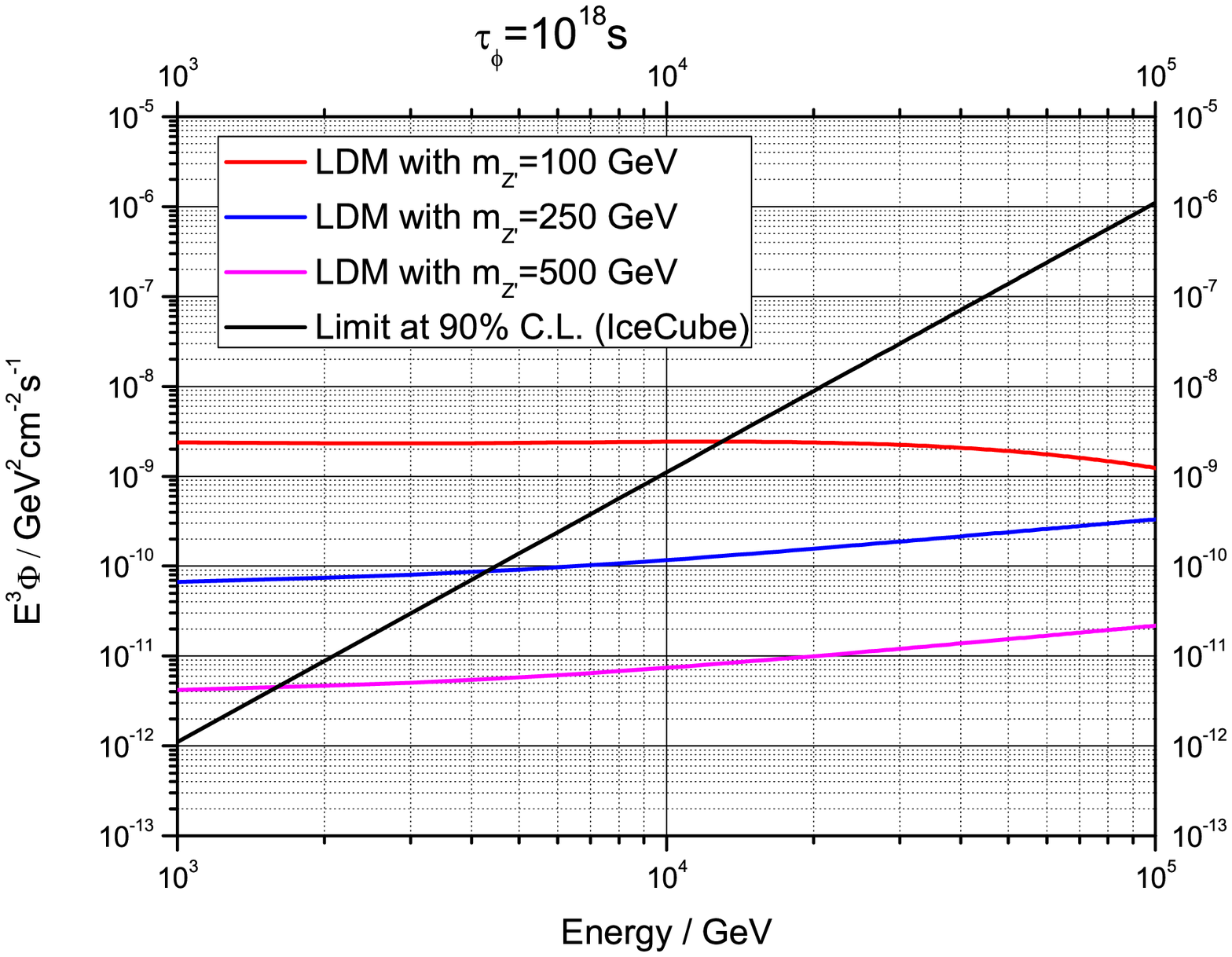}
 \caption{The LDM fluxes and their upper limit at 90\% C.L. are estimated at IceCube with the different $Z^{\prime}$ masses (100 GeV, 250 GeV and 500 GeV) and $\tau_{\phi}=10^{18}$ s, respectively.}
 \label{fig:results3}
\end{figure}


\begin{thebibliography}{}
\bibitem{bergstrom}L. Bergstrom, Rept. Prog. Phys. 63,793 (2000) arXiv: hep-ph/0002126
\bibitem{BHS}G. Bertone, D. Hooper and J.Silk, Phys. Rep. 405, 279 (2005) arXiv: hep-ph/0404175
\bibitem{Planck2015}P.A.R. Abe, et al., Planck collaboration, A\&A 594, A13 (2015) arXiv:1502.01589
%\bibitem{NWF}J. F. Navarro, C. S. Frenk and S. D. M. White, Astrophys. J. 462, 563 (1996) arXiv: astro-ph/9508025
%\bibitem{springel}V. Springel, S. D. M. White, A. Jenkins, C. S. Frenk, N. Yoshida, L. Gao, J. Navarro and R. Thacker et al., Nature 435, 629 (2005) arXiv: astro-ph/0504097
%\bibitem{VLHMR}M. Viel, J. Lesgourgues, M. G. Haehnelt, S. Matarrese and A. Riotto, Phys. Rev. D 71, 063534 (2005) arXiv: astro-ph/0501562
%\bibitem{GK}K. Griest and M. Kamionkowski, Phys. Rev. Lett. 64, 615 (1990)
%\bibitem{GDJ} G.Bertone, D.Hooper, and J.Silk, Physics Reports, 405, 279 (2005)
\bibitem{JP}J.D.Lewin, P.F.Smith, Astropart. Phys. 6, 87 (1996)
\bibitem{CDMSII} R.Agnese, et al., SuperCDMS Collaboration, Phys. Rev. D 92, 072003 (2015)
\bibitem{CDEX} K.J.Kang, et al., CDEX Collaboration, Front. Phys. 8(4),412-437 (2013)
\bibitem{XENON1T} E.Aprile, et al., XENON1T Collaboration, Phys. Rev. Lett. 119, 181301 (2017)
\bibitem{LUX} D.S.Akerib, et al., LUX Collaboration, Phys. Rev. Lett. 118, 251302 (2017)
\bibitem{PANDAX} X.Y.Cui, et al., PandaX-II Collaboration, Phys. Rev. Lett. 119, 181302 (2017)
\bibitem{AMS-02} M.Aguilar, et al., AMS Collaboration, Phys. Rev. Lett. 113, 221102 (2014)
\bibitem{DAMPE} G.Ambrosi, et al., DAMPE Collaboration, Nature, 552, 63-66 (2017)
\bibitem{fermi}A.Albert, et al., Fermi-LAT, DES Collaborations, Astrophys. J. 834, 110 (2017)
\bibitem{KC87} M.Yu.Khlopov, V.M.Chechetkin, Sov. J. Part. Nucl 18, 267-288 (1987)
\bibitem{CKR98} D. J. H. Chung, E. W.Kolb, and A.Riotto, Phys.Rev.Lett. 81, 4048, (1998) arXiv: hep-ph/9805473
\bibitem{CKR99} D. J. H. Chung, E. W.Kolb, and A.Riotto, Phys.Rev. D59, 023501 (1999) arXiv: hep-ph/9802238
\bibitem{KT}V. Kuzmin and I. Tkachev, JETP Lett. 68, 271¨C275 (1998) arXiv: hep-ph/9802304
\bibitem{KCR} E. W. Kolb, D. J. Chung, and A. Riotto, WIMPzillas!, hep-ph/9810361
\bibitem{CKRT} D. J. H. Chung, E. W. Kolb, A. Riotto, and I. I. Tkachev, Phys. Rev. D 62, 043508 (2000) arXiv: hep-ph/9910437
\bibitem{CCKR}D. J. Chung, P. Crotty, E. W. Kolb, and A. Riotto,, Phys. Rev. D 64, 043503 (2001) arXiv: hep-ph/0104100
\bibitem{KST} E. W. Kolb, A. Starobinsky, and I. Tkachev, JCAP 0707, 005 (2007) arXiv: hep-th/0702143
\bibitem{CGIT} L.Covi, M.Grefe, A.Ibarra and D.Tran, JCAP 1004, 017 (2010) arXiv: 0912.3521
\bibitem{FKMY} B.Feldstein, A.Kusenko, S.Matsumoto and T. T.Yanagida Phys. Rev. D 88, 015004 (2013) arXiv: 1303.7320
\bibitem{FKM}M. A. Fedderke, E. W. Kolb, and M. Wyman, Phys. Rev. D 91, 063505 (2015) arXiv:1409.1584
\bibitem{AMO}R. Aloisio, S. Matarrese and A. V. Olinto, JCAP, 1508, 024 (2015), arXiv: 1504.01319
\bibitem{EIP} A.Esmaili, A.Ibarra and O.L. Peres, JCAP, 1211, 034 (2012) arXiv:1205.5281
%\bibitem{FG}P. H. Frampton and S. L. Glashow, Phys. Rev. Lett. 44, 1481 (1980)
%\bibitem{hill}C. T. Hill, Nucl. Phys. B 224, 469 (1983)
%\bibitem{EGL}J. R. Ellis, G. B. Gelmini, J. L. Lopez, D. V. Nanopoulos and S. Sarkar, Nucl. Phys. B 373, 399 (1992)
%\bibitem{BKV}V. Berezinsky, M. Kachelriess and A. Vilenkin, Phys. Rev. Lett. 79, 4302 (1997) arXiv: astro-ph/9708217
%\bibitem{ST}S. Sarkar and R. Toldra, Nucl. Phys. B 621, 495 (2002) arXiv: hep-ph/0108098
%\bibitem{BD}C. Barbot and M. Drees, Phys. Lett. B 533, 107 (2002) arXiv: hep-ph/0202072
%\bibitem{ABK}R. Aloisio, V. Berezinsky and M. Kachelriess, Phys. Rev. D 69, 094023 (2004) arXiV: hep-ph/0307279
%\bibitem{EIP} A.Esmaili, A.Ibarra and O.L. Peres, JCAP, 1211, 034 (2012) arXiv:1205.5281
%\bibitem{BLS} Y.Bai, R.Lu and J.Salvado, JHEP, 01, 161 (2016) arXiv:1311.5864
\bibitem{BGG} A.Bhattacharya, R.Gandhi and A.Gupta, JCAP 1503, 027, (2015) arXiv:1407.3280
\bibitem{BGGM} A.Bhattacharya, R.Gandhi, A.Gupta and S.Mukhopadhyay, JCAP 05, 002 (2017) arXiv:1612.02834
\bibitem{xu1} Y.Xu, JCAP 1805, 055 (2018), arXiv: 1801.06781
\bibitem{xu2} Y.Xu, Physics of the Dark Universe 24, 100287 (2019) arXiv: 1811.09793
\bibitem{xu3} Y.Xu, Astrophys. J. 876, 14 (2019) arXiv: 1804.10719
\bibitem{APQ} A.Alves, S.Profumo and F. S.Queiroz, JHEP 1404, 063 (2014) arXiv:1312.5281
\bibitem{Hooper} D.Hooper, Phys. Rev. D 91, 035025 (2015) arXiv:1411.4079
\bibitem{icecube2017} M.G. Aartsen et al., IceCube collaboration, Astrophys. J. 849, 67 (2017) arXiv:1707.03416
\bibitem{antares}A. Albert et al., ANTARES collaboration, Phys. Rev. D 96, 082001 (2017) arXiv:1706.01857
\bibitem{km3net} S. Adrian-Martinez et al., KM3NeT collaboration, Eur. Phys. J. C 76, 54 (2016) arXiv:1510.01561
%\bibitem{AGASA}N.Hayashida, et al., Phys. Rev. Lett., 77, 1000 (1996)
%\bibitem{HiRes}P. Sokolsky, for the HiRess Collaboration, Nucl. Phys. B (Proc. Suppl.) 212-213, 74-78 (2011)
%\bibitem{Auger2015}A.Abe, et al., the Pierre Auger Collaboration, the Pierre Auger Collaboration to the 34th International Cosmic Ray Conference, 30 July - 6 August 2015, The Hague, The Netherlands, arXiv: 1509.03732
%\bibitem{TA}D. Ivanov, for the Telescope Array Collaboration, in Proc. 34th ICRC 2015, The Hague, The Netherlands, PoS ICRC2015, 349 (2015)
\bibitem{MB} K.Murase and J.F.Beacom, JCAP 1210, 043 (2012) arXiv:1206.2595
\bibitem{RKP} C.Rott, K.Kohri and S.C.Park, Phys. Rev. D 92, 023529 (2015) arXiv:1408.4575
\bibitem{KKK}M.Kachelriess, O.E.Kalashev and M.Yu.Kuznetsov, Phys. Rev. D 98, 083016 (2018) arXiv: 1805.04500
%\bibitem{ARZ}A. R. Zentner, Phys. Rev. D80, 063501 (2009), arXiv:0907.3448
\bibitem{BCH}P. Baratella, M. Cirelli, A. Hektor, et al.,JCAP 1403, 053 (2014), arXiv:1312.6408
\bibitem{gould}A. Gould, Astrophys. J. 321, 571 (1987)
\bibitem{KSW}L. M. Krauss, M. Srednicki and F. Wilczek, Phys. Rev. D 33, 2079 (1986)
\bibitem{nauenberg}M. Nauenberg, Phys. Rev. D 36, 1080 (1987)
\bibitem{GS}K. Griest and D. Seckel,  Nucl.Phys. B 283, 681 (1987)
\bibitem{JKK}G. Jungman, M. Kamionkowski and K. Griest, Phys. Rept. 267, 195 (1996)
\bibitem{LE}J.Lundberg and J.Edsjo, Phys.Rev. D 69, 123505 (2004)
\bibitem{BHM} Martin M.Block, Phuoc Ha, Douglas W.McKay, Phys. Rev. D 82, 077302 (2010) arXiv: 1008.4555
\bibitem{icecube2018}M.G.Aartsen, et al., IceCube Collaboration, Phys. Rev. D 98, 062003 (2018), arXiv:1807.01820
\bibitem{icecube2016}M.G.AArtsen, et al., IceCube Collaboration, Astrophysical J. 833,3 (2016), arXiv:1607.08006
\bibitem{SMS}S.I.Sinegovsky, A.D. Morozova and T.S.Sinegovskaya, Phys. Rev. D 91, 063011 (2015), arXiv:1407.3591
\bibitem{icecubedata2017}M.G. Aartsen et al., IceCube Collaboration, Astrophysical J. 835, 151 (2017)
\bibitem{FC}G. J. Feldman, R. D. Cousins, Phys. Rev. D 57, 3873 (1998)
%\bibitem{Auger2010} J. Abraham, et al., The Pierre Auger Collaboration, Phys. Lett. B 685, 239-246 (2010) arXiv: 1002.1975
%\bibitem{icecube} M.G.AArtsen, et al. , IceCube Collaboration, Astrophysical J. 833,3 (2016) arXiv:1607.08006
%\bibitem{GQRS}R.Gandhi, et al., Astropart. Phys. 5, 81-110 (1996)
%\bibitem{Auger2019}A.Abe, et al., the Pierre Auger Collaboration, Contributions of the Pierre Auger Collaboration to the 36th International Cosmic Ray Conference (ICRC 2019), 24 July - 1 August 2019, Madison, Wisconsin, USA, arXiv: 1909.09073
%\bibitem{ABGPS}R. Aloisio, D. Boncioli, A.F. Grillo, S. Petrera and F. Salamida, JCAP 1210, 007 (2012) arXiv: 1204.2970
%\bibitem{Auger}A.Aab, et al., the Pierre Auger Collaboration, JCAP 1704, 038 (2017) arXiv: 1612.07155
%\bibitem{JEM-EUSO} J.H.Adams, et al. , JEM-EUSO Collaboration, arXiv:1204.5065

%\bibitem{gandhi}Gandhi et al., Phys. Rev. D, 58, 093009 (1998)
%\bibitem{BKMTZ} P.S.Dev Bhupal, et al. JCAP 1608, 034, (2016) arXiv:1606.04517
%\bibitem{eusomission} Y.Takahashi and the JEM-EUSO Collaboration, New J. Phys. 11, 065009 (2009)
%\bibitem{bbfs} M.Bertaina, P.Bobik, F.Fenu , et al., the JEM-EUSO Collaboration, Exp. Astron. 40, 117-134 (2015)
%\bibitem{corsika} D.Heck, et al. 1998, CORSIKA: a Monte Carlo code to simulate extensive air showers, Forschungszentrum Karlsruhe GmbH, Karlsruhe, V + 90 p., TIB Hannover, D-30167 Hannover
%\bibitem{threshold} F.Fenu, for the JEM-EUSO Collaboration , The JEM-EUSO program, arXiv: 1703.01875
%\bibitem[Xu2 2018]{xu2}Xu Y., Directly search for Ultra-High Energy WIMPs at IceCube, arXiv: 1811.09793
%\bibitem{icecube2018}M.G.Aartsen, et al., IceCube Collaboration, Phys. Rev. D 98, 062003 (2018) arXiv:1807.01820
%\bibitem{anita} Derek B.Fox, et al., The ANITA Anomalous Events as Signatures of a Beyond Standard Model Particle, and Supporting Observations from IceCube, arXiv: 1809.09615
%\bibitem[Albert, et al. 2017]{antares}ANTARES collaboration, Albert, A., et al. 2017, Phys. Rev. D 96£¬082001, arXiv:1706.01857
%\bibitem[Adrian-Martinez, et al. 2016]{km3net} KM3NeT collaboration, Adrian-Martinez S. et al. 2016£¬ Eur. Phys. J. C 76, 54, arXiv:1510.01561
%\bibitem[Ichiki,Oguri \& Takahashi 2004]{IOT} Ichiki, K., Oguri, M., and Takahashi, K. 2004, Phys.Rev.Lett. 93, 071302, arXiv:astro-ph/0403164
%\bibitem[Nollett \& Steigman 2015]{NS}Nollett, K. M. and Steigman, G. 2015, Phys. Rev. D 91, 083505, arXiv:1411.6005
%\bibitem[Bhupal,Mazumdar \& Qutub 2014]{DMQ}Bhupal Dev, P., Mazumdar, A. and Qutub, S. 2014, Physics 2, 26, arXiv:1311.5297
%\bibitem[popolo 2007]{popolo}Popolo, A. D. 2007, Astron.Rep. 51, 169¨C196, arXiv:0801.1091
\end{thebibliography}
\end{document}